\documentclass[mathleft
]{an}
\usepackage{graphics,graphicx}
\usepackage{epsf}
\usepackage{epsfig}
\usepackage{color}
\usepackage{times}
\usepackage{natbib}
\bibpunct{(}{)}{;}{a}{}{,}

\newcommand{\Rsun}{\mbox{$R_\odot$}}

\newcommand{\xmm}{{\em XMM-Newton}}

\newcommand{\cxo}{{\em Chandra}}
\newcommand{\hii}{H\,{\sc ii}}
\newcommand{\xicma}{\mbox{$\xi^1$\,CMa}}
\newcommand{\thori}{\mbox{$\theta^1$\,Ori\,C}}
\newcommand{\zcas}{\mbox{$\zeta$\,Cas}}
\newcommand{\voph}{\mbox{V2052\,Oph}}
\newcommand{\bcep}{\mbox{$\beta$\,Cep}}
\newcommand{\tsco}{\mbox{$\tau$\,Sco}}
\newcommand{\zori}{\mbox{$\zeta$\,Ori}}

\newcommand{\Lbol}{\mbox{$L_{\rm bol}$}}
\newcommand{\Tx}{\mbox{$T_{\rm X}$}}
\newcommand{\Lx}{\mbox{$L_{\rm X}$}}

\newcommand{\Msun}{\mbox{$M_\odot$}}
\newcommand{\myr}{\mbox{$M_\odot\,{\rm yr}^{-1}$}}
\newcommand{\lsim}{\raisebox{-.4ex}{$\stackrel{<}{\scriptstyle \sim}$}}
\newcommand{\msim}{\raisebox{-.4ex}{$\stackrel{>}{\scriptstyle \sim}$}}

\def \etal   {\hbox{et~al.\/}}
\begin{document}

\Pagespan{789}{}
\Yearpublication{2006}%
\Yearsubmission{2005}%
\Month{11}%
\Volume{999}%
\Issue{88}%

\title{X-ray emission from massive stars with magnetic fields
\thanks{Based on observations obtained with
\xmm\ and \cxo}}

\author{L. M. Oskinova\inst{1}\fnmsep\thanks
{Corresponding author: \email{lida@astro.physik.uni-potsdam.de}\newline}
\and W.-R.\,Hamann\inst{1}
\and J.\,P.\, Cassinelli\inst{2} 
\and J.\,C.\,Brown\inst{3} 
\and H. Todt\inst{1} 
}  
\titlerunning{X-rays from massive stars with magnetic fields}
\authorrunning{L.M. Oskinova \etal.}
\institute{
Institute for Physics and Astronomy, University of Potsdam,
14476 Potsdam, Germany 
\and
Department of Astronomy, University of Wisconsin-Madison, Madison, WI 53711, 
USA
\and
School of Physics and Astronomy, University of Glasgow, Glasgow G12 8QQ, UK
}
\received{30 May 2005}
\accepted{11 Nov 2005}
\publonline{later}


\abstract{We investigate the connections between the magnetic fields
  and the X-ray emission from massive stars.  Our study shows that the
  X-ray properties of known strongly magnetic stars are diverse: while
  some comply to the predictions of the magnetically confined wind
  model, others do not. We conclude that strong, hard, and variable
  X-ray emission may be a sufficient attribute of magnetic massive
  stars, but it is not a necessary one. We address the general
  properties of X-ray emission from ``normal'' massive stars,
  especially the long standing mystery about the correlations between
  the parameters of X-ray emission and fundamental stellar
  properties. The recent development in stellar structure modeling
  shows that small scale surface magnetic fields may be common.  We
  suggest a  ``hybrid'' scenario which could explain the X-ray
  emission from massive stars by a combination of magnetic mechanisms
  on the surface and shocks in the stellar wind. The magnetic
  mechanisms and the wind shocks are triggered by convective motions
  in sub-photospheric layers.  This scenario opens the door for a
  natural explanation of the well established correlation between
  bolometric and X-ray luminosities.}

\maketitle
\section{Introduction}

Massive stars $(M_{\rm initial}\geq 10 M_\odot)$ are among the key
players in the cosmic evolution.  Early UV observations revealed that
their outer envelopes are outflowing in the form of powerful stellar
winds \citep{mor67}. The winds of OB-type and Wolf-Rayet (WR) type
stars are driven by radiati\-ve \ pre\-ssure on metal lines (CAK,
Castor \etal\ 1975, Gr{\"a}fener \& Hamann 2005). In general, \ the
wind power \ depends on the evolutionary \ status of the star and is
strongest for the \ evolved WR-type stars.  There are numerous
observational and theoretical evidences that magnetic fields play an
important role in inner as well as in outer layers of massive
stars. Magnetic fields in massive stars have the potential to strongly
influence stellar formation and evolution \citep[e.g.][]{fer09} and
affect stellar winds \citep[e.g.][]{bm97}.

While direct measurements of magnetic fields have so far been possible
only for the closest and brightest stars, the indirect evidence for
their presence is wide spread.  The wide range of observational
phenomena, such as wind-line periodic variability
\citep[e.g.][]{ham01} and excess emission in UV-wind lines centered
about the rest wavelength, are commonly explained by the influence
that magnetic fields exert on stellar winds
\citep[e.g.][]{sch08}. Chemical peculiarity, specific pulsation
behavior, and non-ther\-mal radio emission may be manifestations of
magnetic fields in massive stars as well as their X-ray emission.

In this paper we consider the X-ray emission from \ massive \ stars and
its possible association \ with magnetic phenomena.\ In Section \ 1 we
address the \ ``normal'' mas\-sive stars, \ and put \ forward a hybrid
scenario for their X-ray emission; in Section\,2 we consider stars where
magnetic fields have been directly measured; in Section\,3 the influence
of the X-ray emission on stellar winds is briefly discussed. In
Section 5 we finally consider the X-rays from O-stars on the zero-age
main sequence and from WR-stars; concluding remarks are given in
Section\,6.

\section{Magnetic fields can be important to understand the  X-rays
from ``normal'' massive stars}

Stars across the HR diagram emit X-rays. Low- and solar-mass stars
possess X-ray emitting coronae which are powered by an outer convection
zone via magnetic fields. Stars of spectral types earlier than
$\sim$A7 have no outer convection zone and normally no surface magnetic
field.  This text-book picture was challenged by works of
\citet{maccas03} and most recently by \citet{cb2011}.  They found that
magnetic fields  of sufficient amplitude to affect the wind could emerge
at the surface  via magnetic buoyancy and suggested that this type of
surface  magnetism could be responsible for photometric variability and
play a role in the generation of the X-ray emission and wind clumping.

These theoretical insights in stellar structure are \ in a good
agreement with \ what is known about \ X-ray emission from massive
stars. \ Already first X-ray observations of O-stars by the {\em
  Einstein} observatory revealed the emission from high ions like
S\,{\sc xv} and Si\,{\sc xiii} in the O9.7Ib star \zori\ \citep{cs83}.
Interestingly, \citet{bs96} reported an X-ray flare from \zori. The
analysis of the high-resolution spectrum of \zori\ revealed that the
hottest plasma is located close to the stellar core and that the wind
is quite transparent for X-rays \citep{wc01}.  So\-on after a surface
magnetic field on \zori\ was detected  by \citet{bour08}. Larger
samples of X-ray spectra \ from \ O-type \ stars \ were \ analyzed in
\citet{leut06,osk06,wc07,raa2008}. All these works agree that the 
hot\-test X-ray emitting plasma can be found within 0.5\,$R_\ast$ 
from the stellar surface while the cooler X-ray emitting gas is spread 
out through the extended stellar wind.

\citet{cs83} suggested that two mechanisms of X-ray emission could
operate in massive stars: very hot, \ probably \ magnetically \ confined
loops \ near the base \ of the wind and \ fragmented shocks \ embedded in 
the wind. \ Since the X-ray variability was already known to be less than
about 1\%, \citet{cs83} suggested that there are thousands of shock
fragments in the wind. This is recently confirmed by newer data of
better quality (Naz\'e \etal\ 2011). Radiation hydrodynamic
simulations of the nonlinear evolution of instabilities in stellar
winds were performed by \citet{ocr88}. They demonstrated that X-ray
emission can originate from plasma heated by strong reverse shocks,
which arise when a high-speed, rarefied flow impacts on slower
material that has been compressed into den\-se shells. \citet{feld97}
showed that in order to match the observed X-ray flux, the wind shocks
must be triggered by an instability seed perturbation at the base of
the stellar wind.

We suggest that these seed perturbations at the base of the stellar
wind could be associated with subsurface convective regions. The
following phenomenological picture of X-ray emission from hot massive
stars is emerging: the subsurface convective regions caused by an
opacity peak associated with iron could host a dynamo, producing
magnetic fields reaching the surface. The surface magnetic structures
could play a pivotal role in generating the hottest plasma observed in
X-ray spectra (an example of such mechanism is considered in
\citet{wc09}). The turbulent motions in the convective zone also
provide the seed perturbations required to trigger sufficiently strong
shocks to match the observed X-ray fluxes. Furthermore, as
hydrodynamic simulations show, these instabilities result in the
structuring of the cool wind into fragmented dense shells
(colloquially speaking ``clumps'').  Such porous wind structure makes
the wind relatively transparent for X-rays \citep{feld03}.

The correlation between stellar bolometric and X-ray luminosity,
$L_{\rm X}\approx 10^{-7}L_{\rm bol}$ is well established (most
recently, \citet{naza11,gag11}), but is not well understood. In our
new scenario the X-ray emission is correlated with the stellar
parameters in a natural way, via the dependence of the properties of
the convective zone on fundamental stellar parameters $L_{\rm bol}$
and $T_{\rm eff}$.  \citet{cb2011} \ predict \ that \ the surface 
\ magnetic field \ strength is increasing \ for hotter and more 
\ luminous stars. The  wind momentum \ is also increasing with luminosity
\citep{puls1996}. Thus, the ratio of the wind kinetic energy to the
magnetic energy could remain constant for stars of different
luminosities. 

The $L_{\rm X}\approx 10^{-7}L_{\rm bol}$ correlations breaks down for
stars with $\log(L_{\rm bol}/L_\odot)\lsim 4.4$ \citep{sana2006}. Such
stars are located within the $\beta$\,Cep-type instability domain in
the HR diagram and have specific a pulsational behavior ex\-plai\-ned
by the $\kappa$-mechanism \citep{dz1993}. The break down of $L_{\rm
  X}\propto L_{\rm bol}$ correlation for $\beta$\,Cep type pulsators
provides further clues on the connection between the X-ray emission
and the stellar structure.

A correlation between the X-ray and the effective stellar temperatures
was found by \citet{wal2009} from their studies of high-resolution X-ray
spectra of a sample of massive stars.  It is interesting to note that in
the time-dependent hydrodynamic simulations by \citet{feld97} the
velocity jump U of the wind shocks depends on the ratio between the
period of the perturbations at the wind base, $T_{\rm c}$, and the flow
time, $T_{\rm flow} = R_\ast/v_\infty$.  According to \citet{cb2011},
the convective turnover time ($\sim$\,hr) is larger for more massive
stars. Since $v_\infty\propto \sqrt{M_\ast R_\ast^{-1}}$, the velocity
jump, i.e.\ the temperatures of the gas heated in the shock could be
higher for more massive stars with higher $T_{\rm eff}$. This could
explain the correlations  between $T_{\rm X}$ and $T_{\rm eff}$
discovered by \citet{wal2009}.

\section{Magnetic fields on massive stars are not necessarily manifested
via strong, hard and variable X-rays}
 
From the small-scale magnetic fields which we considered in previous
section, we now turn to the large-scale organized fields that may be
fossil in origin. \citet{bm97} studied the case of a star with a
stellar wind and a dipole magnetic field. They predicted that a collision
between the wind components from the two hemispheres in the closed
magnetosphere leads to a strong shock and characteristic X-ray emission.
Based on this magnetically confined wind shock model (MCWS), the
presence of a magnetic field on the O-type star $\theta^1$\,Ori\,C had
been postulated. Direct confirmation of the magnetic field in this star
by \citet{don06}  proved that X-rays have large diagnostic potential in
selecting massive  stars with surface magnetic fields.

Using the parameters of $\theta^1$\,Ori\,C, \citet{ud02} and
\citet{gag05}  performed MHD simulations in the framework of the
MCWS model and made predictions that can be directly compared with
observations: {(\it i)} the hottest plasma should be located at a few
stellar radii from the stellar surface at the locus where the wind
streams collide; {(\it ii)} the X-ray emission lines should be rather
narrow, because the hot plasma is nearly stationary; {(\it iii)} magnetic
stars should be more X-ray luminous than their non-magnetic counterparts
of similar spectral type; {(\it iv)} the X-ray spectrum of magnetic stars
should be harder than that of non-magnetic stars, with the bulk of the
hot plasma at the hottest temperature; {(\it v)} the X-ray emission
should be modulated periodically as a consequence of the occultation of
the hot plasma by a cool torus of matter, or by the opaque stellar core.
All these predictions are fulfilled for $\theta^1$\,Ori\,C. This modeling
success established the MCWS model as a general scenario for the X-ray
emission from magnetic early type stars. 

Meanwhile, new observations of X-rays from magnetic O-type stars have
been obtained. A strong magnetic field  ($\sim 1$\,kG) is detected on
HD\,108 (O7I) \citep{mar10}. However, the emission measure (EM) of the
softer spectral component, with a temperature of $\approx 2$\,MK, is
more than one order of magnitude higher than the EM of the harder
component $T_{\rm max}\approx 15$\,MK, contrary to the expectation of
the MCWS model \citep{naze04}.  HD\,191612 also has a $\sim 1$\,kG
strong magnetic field (Donati \etal\ 2006a).  \citet{naze10}
demonstrated that the large EM at $\approx$2\,MK and the broad X-ray
emission lines of this star do not compare well
with the predictions of the MCWS model. Overall, considering the
analysis of X-ray observations of magnetic O stars, it appears that only
one star, $\theta^1$\,Ori\,C, displays  properties that are fully
compatible with the predictions of the MCWS model.

\citet{cas02} and \citet{jcb08} studied the case of fast rotating magnetic massive
stars, specifically addressing the formation of disks in classical
Be-type stars. They showed that magnetic torquing and channeling of wind
flow from intermediate latitudes of the stellar surface can, for
plausible field strengths, create a dense disk a few stellar radii in
extent.  \citet{li08} proposed a model, where the X-rays are produced
by wind material that enters the shocks above and below the disk
region.  The model by Li \etal\ predicts a relation
between the X-ray luminosity normalized to the stellar bolometric
luminosity (\Lx/\Lbol) and the magnetic field strength in Be-type stars.

An interesting and well-studied example of a star with quite strong
surface magnetic field is \tsco.  It displays several unusual features:
(1) redshifted absorption in UV P Cyg lines to approximately
+250\,km\,s$^{-1}$ suggestive of infalling gas, (2) unusually hard X-ray
emission requiring hot plasma at temperatures in excess of 10\,MK
\citep{ws05,mewe03}, (3) a complex photospheric magnetic field with open
and closed field lines \citep{dtsco06}. 

\begin{figure}[h]
\centering
\includegraphics[width=0.9\columnwidth]{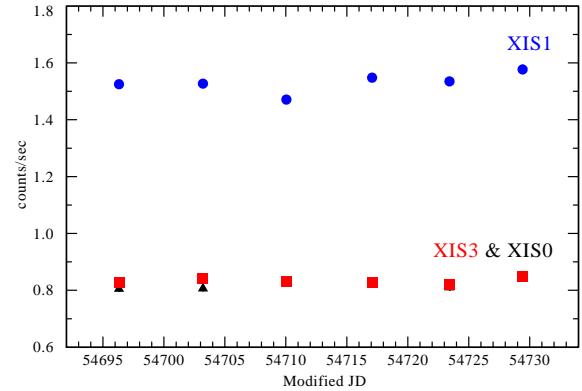}
\caption{The {\em Suzaku} X-ray lightcurve of \tsco. The upper curve is
  observations by the XIS1 detector in 0.2--12.0\,keV band. The lower
  curves are observations with the XIS0 and XIS3 detectors (0.4--12\,keV
  band).  The error-bars are smaller than
  the size of the symbols.}
\label{fig:tscolc}
\end{figure}

We obtained six observations of \tsco\ with the {\em Suzaku} X-ray
observatory to roughly sample its rotational period.  The result of
these observations  were quite surprising \citep{ig10}.  No modulation
of the X-ray emission on a level above 3\,\%\ was detected (see
Fig.\,\ref{fig:tscolc}), while the MCWS model predicts 40\%
\citep{dtsco06}. It appears that the spatial distribution of the hot gas
in \tsco\ is different from the large-scale magnetic field distribution.

Donati \etal\ noted that the absence of the \ predicted \ high-amplitude
X-ray \ modulation \ could be an indicator \ of smal\-ler scale magnetic \ 
loops and confined hot gas across the stellar surface. Such loops would have
evaded detection in their study. It is important to note that  X-ray
variability has not been detected from another magnetic B-star \bcep\
\citep{fav09}. 

In \citet{osk11} we \ investigat\-ed the \ X-ray em\-ission and wind
properties of \ magnetic B-type stars. We studied all magnetic stars
earlier than B2 with available X-ray data.  Dedicated
observations with \xmm\ were performed for the three \ magnetic \ B-stars
\xicma, \ \voph, \ and \zcas\ \citep{osk11}.  Two of them, \voph\ and \zcas, 
\ are detected
in X-rays for the first time. We also searched the X-ray archives and
collected the X-ray data for other magnetic early B-type stars.

One of the key findings of this study was that some magnetic early
type stars have soft X-ray spectra. An interesting example is
\zcas. This star is an oblique magnetic dipole with polar field
strength $\approx 335$\,G \citep{nzcas2003}.

\begin{figure}
\centering
\includegraphics[height=0.99\columnwidth, angle=-90]{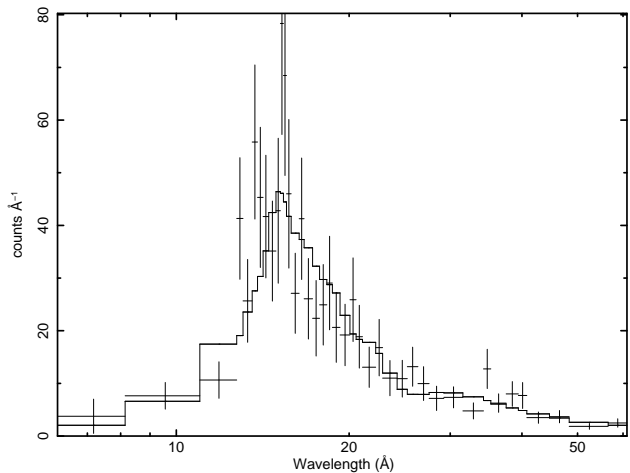}
\caption{ \xmm\ EPIC PN 
spectrum of \zcas\ and the best fit two-temperature model.}
\label{fig:zcassp}
\end{figure}

Our \xmm\ observations of \zcas\ is the first to detect X-rays from this
star.   The EPIC spectra of \zcas\ and the best fit two-temperature
model are shown in Fig.\,\ref{fig:zcassp}.  The emission measure is
dominated by plasma of 1\,MK; a hotter, 4\,MK component constitutes
less than 20\% of the total emission measure.  There are no indications
of a harder spectral component, making  \zcas\ the \ softest \ X-ray
 source \ among all hot stars \ where magnetic field \ have been detected.
\ The mean
X-ray \ spectral \ temperature of \zcas\ is about 1\,MK and its X-ray
luminosity is $\approx 5\times 10^{29}$\,erg\,s$^{-1}$  \citep{osk11}.

Our study of magnetic early B-type stars revealed that the X-ray
spectra and fluxes of early B-type stars with confirmed magnetic
fields do not significantly differ from the X-ray spectra of stars
where magnetic fields have not been found
\citep[e.g][]{raa2005}. Similar conclusions were reached by
\citet{pet10}. It implies that either magnetic fields play only a minor
role in the X-ray generation, or that magnetic fields are present
(but remained yet undetected) in all early B-type stars emitting X-rays.

%

The X-ray emission from the peculiar magnetic Bp-type stars is diverse
\citep{drake1994,osk11}. While some stars display hard variable
X-rays, others are rather soft sources. The X-ray luminosities differ
among the otherwise similar stars in this group by more than two
orders of magnitude.

We must conclude that the role of magnetism in the generation of X-ray
emission in stellar winds, and consequently the physics of winds in
magnetic stars, is not fully understood. New observational data and
new MHD modeling of stellar winds shall bring future
insights.

\section{Magnetic fields and the X-ray emission strongly affect stellar
winds}

Information about stellar winds from B stars with magnetic fields
was obtained by \citet{osk11} by modeling of the UV spectra.
We found that the mass-loss rate in \zcas\ does not exceed
$10^{-9.7}$\,\myr. This is $\approx 8$ times smaller than the prediction
of \citet{bab1996}. Furthermore, we did not find any evidence for the
predicted fast wind velocity of $v_{\infty} = 2100$\,km\,s$^{-1}$. 
Similarly, for all other stars in our sample, the mass-loss rates are an
order of magnitude lower than predictions by \citet{abb1982} based on
the CAK theory.  

We find that, although the X-rays strongly affect the ionization
structure of the wind, this effect is not sufficient in reducing the
total radiative acceleration. When the X-rays are accounted for at the
intensity and temperatures observed, there is still sufficient radiative
acceleration to drive stronger mass-loss than we empirically infer from
the UV spectral lines.

The emission measure of the hot X-ray emit\-ting gas
sig\-nifi\-cant\-ly (by 3-4 orders of magnitude) exceeds the emission
measure of the cool wind, where the UV and optical spectra
originate. This is an old standing problem, first pointed out by
\citet{cass1994,coh1997}.  Our new analyses of better quality X-ray
and optical and UV spectra further aggravate this problem.

\section{Magnetic fields may be present on very young massive stars and
on very old massive stars}

X-ray observations of massive stars have the potential to probe
magnetic fi\-elds at different evolutionary stages.  The growing
number of sensitive X-ray observations of young star-forming regions
(SFRs) allowed the detection and study of recently formed massive
stars.  In presence of a strong surface magnetic field, a higher X-ray
luminosity combined with a hard, sometimes non-thermal, spectrum, and
X-ray variability may be expected. {\em Chandra} observations 
revealed that very young massive stars in the Galactic complex W3 are
emitters of hard X-rays ($\sim$7~keV). \citet{hof02} propose that
magnetic reconnection events can be a mechanism responsible for this
emission. However, they point out that the expected X-ray variability
is detected only in one out of ten massive young stars.  Magnetic
fields have also been proposed to explain \ the hard \ X-ray spectrum
($kT_{\rm X}\msim$2~keV) and \ synchrotron X-ray \ emission from the
vicinity of the O star IRS\,2, \ the primary source of ionization of the
\hii\ region RCW\,38 \citep{wolk02}. The template magnetic massive
star \thori\ is also located at the heart of a very young massive star
cluster. Motivated by these observational findings, \citet{sch03}
proposed that the strength of stellar surface magnetic fields decline
with stellar age.

\begin{table*} 
\caption{X-ray properties of O stars in some star clusters younger 
than  2\,Myr} 
\vspace{1mm} 
\center 
\begin{tabular}{lcccccc} 
\hline 
Name & Age & Sp.Type &$\log(\Lx/\Lbol)$ & Presence of hard emission &  
$\langle\log(\Lx/\Lbol)\rangle$ & Ref.  
\rule[-1mm]{0mm}{4.5mm}\\  
&Myr & earliest O-type star&earliest O-type star&$k\Tx\msim 2$\,keV &  
ensemble of O stars & \\ \hline 
\rule[-1mm]{0mm}{4.5mm} 
W3         & 0.2      &05-6V        & $\sim -6.7$   & yes &            & 1\\ 
Trifid     & 0.3      &O7.5III      & $\lsim -7.2$  & no  & $\lsim -7$ & 2\\ 
RCW\,38    & $\lsim 1$&05V          & -5.3          & yes & -5.6       & 3\\ 
NGC\,3603  & 1        &03-4V        & $\sim -6$     & yes & -6..-8     & 4\\ 
NGC\,6618  & 1        &O4V+O4V      & -6.7          & yes &            & 5\\  
Trapezium  & 1        &O6V          & -5.8          & yes & -6.5       & 6 \\ 
Hourglass  & 1        &O7V          & -8            & no  &            & 5\\ 
SMC N81    & 1        &O6.5V        & -             & -   & $\lsim -7$ & \\ 
NGC\,6611  & 1.3      &O5V          & -6.8          & no  &            &
5,7\\	       
Rosette    & 1.9      &04V          & -7            & no  & $\sim -7$  & 5\\ 
\hline 
\rule[-1mm]{0mm}{1.5mm}\\  
\multicolumn{7}{l}%
{\footnotesize  
(1) \citet{hof02}; (2) \citet{rho04};  
(3) \citet{wolk06}; (4) \citet{mof02};} \\ 
\multicolumn{7}{l}%
{\footnotesize  
(5) \Lx\ is inferred using archival {\em Chandra} data;  
(6) \citet{feig02}; 
(7) Cluster parameters from  from \citet{bon06};  } \\ 
\end{tabular} 
\vspace{-0.3cm} 
\label{tab:cl} 
\end{table*} 

To check whether the X-ray activity of massive stars is especially
strong in youngest stars, we compiled in Table~\ref{tab:cl} a list of
young ($\lsim$2~Myr) clusters observed by {\em Chandra}. We ha\-ve
inferred the X-ray luminosity of O stars in NGC\,6618, NGC\,6611, and
the Hourglass and Rosette Nebulae using {\em Chandra} ar\-chival data.
X-ray observations of O stars in the other clusters in
Table\,\ref{tab:cl} have been taken from the literature. Using the
typical bolometric luminosity for each spectral type \citep{mar}, we
list in Table\,\ref{tab:cl} the \Lx/\Lbol\ ratio for the earliest O star
in each cluster. We also indicate whether the X-ray spectra are
``hard''. In cases when analyses of the X-ray emission from the whole
cluster population of O stars are available, the average ratio
$\langle\log(\Lx/\Lbol)\rangle$ is given.

As Table\,\ref{tab:cl} illustrates, the X-ray activity of massive
stars differs significantly from cluster to cluster.  Profoundly, in
RCW\,38 and the Orion Trapezium, the earliest O stars are more active
than in others clusters of similar age. On the other hand, some stars
are weak X-ray sources, e.g., Her\,36 (O7V) in the Hourglass and
HD\,164492A (O7.5III) in the Trifid. Her\,36 is of similar age and
spectral type as \thori, yet the latter is significantly more X-ray
luminous.

The hardness of the X-ray spectrum is not an unambiguous indication of
the presence of a magnetic field. X-ray temperatures of several keV
are expected and have been observed in binary systems. The probability
that a massive star in a young cluster is a binary is high, therefore
it is not surprising that many stars listed in Table\,\ref{tab:cl}
display the presence of $k\Tx\msim$2~keV plasma in their spectra.
This is likely the case of the Kleimann star, a massive O4+O4 binary
ionizing the Omega Nebula cluster NGC\,6618.  While its spectrum is
relatively hard, the broad-band X-ray luminosity is consistent with
the canonical $\Lx\approx 10^{-7} \Lbol$\ relation which holds also
for binaries (Oskinova 2005).

The X-ray observations of the Trifid Nebula \citep{rho04}
allowed to resolve the central ionizing source into discrete components 
and revealed that an O7.5III star has a soft ($kT\approx 0.6$\,keV)  
spectrum, while har\-der emis\-sion ($kT\approx$6~keV) is associated with  
a B-type star which is blen\-ded with an unidentified source.
Interestingly,  X-ray fla\-res were previously detected from B-type
stars in two  Orionis clusters: Trapezium and $\sigma$\,Ori 
\citep{sf04,st05}.  

Often, young SFRs where massive stars are just forming are located
close to more evolved massive star clusters, suggesting a possible
causal connection between massive star feedback and star formation.
An example of such SFR is ON\,2 located close to the massive star
cluster Berkeley\,87. Its X-ray properties were explored in Oskinova
\etal\ (2010).

The most evolved star in Berkeley\,87 is a rare WO-type star, WR\,142. 
Stars of this spectral type are the ultimately latest evolutionary stage
of a very massive stars \citep[e.g.][]{san2011}.  There are no direct
measurements of magnetic fields on WR stars up to now (see Kholtygin
\etal, these proceed.).  However, the X-ray properties of WR stars may
provide indirect evidence that magnetism plays a role also in the latest
stages of stellar evolution.

Our \xmm\ observation of the closest WO-type star WR\,142 succeeded in
detecting this object \citep{osk09}. It was also detected by the  {\em
Chandra} observatory \citep{sok2010}. Albeit the signal-to-noise of the
observed spectrum was too poor for a detailed spectral analysis, from
the hardness ratio \citet{osk09} concluded that the X-ray emission from
this star is strongly absorbed in the wind,  and is too hard to be
explained by the wind-shock mechanism. We speculated that the
hypothetical magnetic magnetic field can be responsible for this strong
emission.

Our analysis of the optical spectrum of this star yielded the
fundamental stellar parameters: $R_\ast = 0.5$\Rsun, $T_{\rm
  eff}=160$\,kK, $M_{\ast}=5\Msun$. \ Remarkedly, the optical emission
line profiles in the spectrum of WR\,142 \  have an \ unusual round
shape. The \ formal rotational \  broadening models of \ these lines yield
projected \ rotational \ velocity $v\sin{\rm i}=4000$\,km\,s$^{-1}$.
Thus, the star may be rotating at break-up velocity. These stellar
parameters compare well with those expected for a SN or $\gamma$-ray
burst progenitor \citep{pacz98}.

\section{Concluding remarks} 

Surface magnetic fields on massive stars are both predicted
theoretically and confirmed observationally. All \ assive stars emit
X-ray radiation, but the \ connection between X-rays and magnetism in
massive stars is not yet fully understood. The MCWS model can 
explain the strong, hard, varable X-ray emission from some magnetic
dipoles. However there is no quantitative explaination for the
soft and the constant X-ray emission that is observed in the majority
of magnetic massive stars.

It was recently argued that {\em all} massive stars may have magnetic
fields on their \ surface \ \citep{cb2011}. In this paper we propose a
new scenario of X-ray emission from a massive star, where the plasma  on
the surface is heated  by magnetic mechanisms while the plasma embedded
in the stellar wind is heated by  wind shocks. Both these mechanisms are
triggered by sub-photospheric convective motions.

The advance of spectropolarimetric and  X-ray observational
techniques will undoubtedly lead to a new understanding of massive
stars physics.

\acknowledgements 
This research has made use of NASA's Astrophysics Data
System Service and the SIMBAD database, operated at CDS, Strasbourg,
France.  Funding for this research has been provided by DLR grant 50\,OR\,0804 
(LMO) and a UK STFC Grant (JCB)+ .    


\end{document}